%% file: main.tex
\documentclass[
    reprint,
    nofootinbib,
    aps,
    prd,
    amsmath,
    floatfix,
    ]{revtex4-2}
\usepackage{accents}
\usepackage{amsfonts}
\usepackage{bm}
\usepackage{graphicx}
\usepackage[colorlinks, allcolors=blue]{hyperref}
\usepackage{xcolor}

\definecolor{ShamrockGreen}{rgb}{0.0, 0.62, 0.38}

\begin{document}

\title{TopicFlow: Disentangling quark and gluon jets with normalizing flows}

\author{Matthew J.\ Dolan}
\email{matthew.dolan@unimelb.edu.au}
\affiliation{ARC Centre of Excellence for Dark Matter Particle Physics, School of Physics, The University of Melbourne, Victoria 3010, Australia}
\author{Ayodele Ore}
\email{ayodeleo@student.unimelb.edu.au}
\affiliation{ARC Centre of Excellence for Dark Matter Particle Physics, School of Physics, The University of Melbourne, Victoria 3010, Australia}

\date{\today}
\begin{abstract}
The isolation of pure samples of quark and gluon jets is of key interest at hadron colliders. Recent work has employed topic modeling to disentangle the underlying distributions in mixed samples obtained from experiments. However, current implementations do not scale to high-dimensional observables as they rely on binning the data. In this work we introduce TopicFlow, a method based on normalizing flows to learn quark and gluon jet topic distributions from mixed datasets. These networks are as performant as the histogram-based approach, but since they are unbinned, they are efficient even in high dimension. The models can also be oversampled to alleviate the statistical limitations of histograms. As an example use case, we demonstrate how our models can improve the calibration accuracy of a classifier. Finally, we discuss how the flow likelihoods can be used to perform outlier-robust quark/gluon classification.
\end{abstract}

\maketitle
\section{Introduction}

The primary objects of study in hadronic physics at the Large Hadron Collider (LHC) are jets. A fundamental probabilistic question that particle physicists often ask regarding jets is: what is the parton-level of origin of a given jet? Unfortunately for the question of quark versus gluon jet-tagging, there is no clear hadron-level definition of the distinction between a quark-jet and a gluon-jet~\cite{Banfi:2006hf, Gras:2017jty}. While one can nonetheless train classifiers based on signal and background samples with ``truth" parton-level information obtained from Monte-Carlo simulation \cite{Gallicchio:2011xq,Komiske:2016rsd,Cheng:2017rdo,Kasieczka:2018lwf,Lee:2019cad,Dreyer:2021hhr,Romero:2021qlf,Bright-Thonney:2022xkx}, these definitions are subject to large modeling uncertainty \cite{Reichelt:2017hts, Mo:2017gzp}. As such, recent research has explored the utility of operational definitions. These are often based on data-driven mixture models, such as jet topics~\cite{Metodiev:2018ftz,Komiske:2018vkc,Komiske:2022vxg}, Latent Dirichlet Allocation (LDA)~\cite{Dillon:2019cqt,Dillon:2020quc,Dillon:2021aeo} and others~\cite{Alvarez:2021zje}. 
Using these methods one can disentangle samples containing multiple underlying components, with applications including anomaly detection, weakly-supervised classification and studies of quantum chromodynamics (QCD). As these definitions are operational, they also have the advantage of avoiding issues such as systematic errors in the modeling of QCD and detector effects. On the other hand the extracted quark and gluon distributions may change when different algorithms are used, motivating the development of further techniques for disentangling quark and gluon distributions in LHC data.

To date, data-driven disentangling has been performed primarily in single observables, such as jet multiplicity. While jet topics and LDA technically generalize to higher dimension, their dependence on binning causes increased computational cost and statistical error in this regime. On the other hand, deep generative networks can model probability distributions in an unbinned manner and have not yet been explored for topic modeling in jet physics. Normalizing flows are examples of such generative models and have proven to be a powerful tool for many tasks in collider physics, including speeding up various parts of the event generation pipeline \cite{Stienen:2020gns,Bothmann:2020ywa,Gao:2020vdv,Gao:2020zvv,Lu:2020npg, Butter:2021csz,Krause:2021ilc,Krause:2021wez,Kach:2022qnf,Kach:2022uzq,Verheyen:2022tov}, detecting anomalous events \cite{Choi:2020bnf,Nachman:2020lpy,Hallin:2021wme,Jawahar:2021vyu,Butter:2022lkf, Hallin:2022eoq} and more \cite{Bellagente:2020piv, Bieringer:2020tnw, Winterhalder:2021ngy, Winterhalder:2021ave, Butter:2022vkj,Klein:2022hdv}. Recent reviews of LHC event generation in the context of machine learning are Refs.\,\cite{Butter:2020tvl,Butter:2022rso}.

In this work, we build on the jet topics framework by leveraging normalizing flows. We train models to fit the topic distributions directly from mixed datasets, eliminating the need to subtract histograms which compounds statistical uncertainty. The benefits of our approach are therefore twofold. First, since the flows are unbinned, they can be applied efficiently in large dimension which allows high-order correlations to be captured. This opens the door for combined use with other machine learning models that are sensitive to such correlations. Second, the flows can be oversampled to obtain smooth distributions, mitigating statistical error. This is particularly relevant for high purity quark/gluon mixtures, which can only be obtained with small cross-section~\cite{Gallicchio:2011xc}. An additional feature of the training procedure is that it explicitly encourages separation of the classes in the latent space. This can be exploited to define an outlier-robust classifier using the likelihood ratio defined by two learned topic distributions.

In the remainder of this section, we provide background on the jet topics framework before motivating a generative approach and describing the training procedure for our normalizing flows. Section~\ref{sec:training} details our specific datasets, network architecture and training parameters. In Section~\ref{sec:results} we demonstrate the quality with which our normalizing flows can distill underlying jet topic distributions, comparing to the histogram-based construction. We also present two use cases: improving data-driven classifier calibration and performing generative classification. We provide concluding remarks and discuss further applications of these models in Section~\ref{sec:conclusions}.

\subsection{Jet topics}
We consider samples of jets as unlabelled statistical mixtures of quarks and gluons. Assuming that two mixtures~$M_1$ and $M_2$ consist of identical quark and gluon component distributions $p_Q$ and $p_G$, the probability densities for an observable $x$ can be written as
\begin{align}
    p_{M_1}(x) &= f_1 \, p_Q(x) + (1-f_1) \, p_G(x) \notag\\
    p_{M_2}(x) &= f_2 \, p_Q(x) + (1-f_2) \, p_G(x),
\end{align}
with $f_1$ and $f_2$ being the unknown quark fractions of the mixtures. In the jet topics framework~\cite{Metodiev:2018ftz, Komiske:2018vkc, Komiske:2022vxg}, one further assumes that the component distributions are \emph{mutually irreducible}, meaning that there exist regions of phase space which can be identified as being either pure-quark or pure-gluon. The mixtures can be disentangled using the \textsc{Demix} algorithm~\cite{pmlr-v33-blanchard14}, which proceeds by calculating \emph{reducibility factors} defined as
\begin{align}
    \label{eq:red-factors}
    \kappa_{ij} = \underset{x}{\operatorname{min}}\,\frac{p_{M_i}(x)}{p_{M_j}(x)}\,,
\end{align}
for $i,j\in\{1,2\}$. These factors correspond to the maximal amount by which $p_{M_j}$ can be subtracted from $p_{M_i}$ while ensuring the result is positive-definite. Ref.\,\cite{Komiske:2022vxg} provides a number of methods for calculating these factors, the simplest of which amounts to binning the mixtures in the observable $x$ and performing the minimization of Eq.\,\ref{eq:red-factors} over these bins. The underlying ``topic" distributions may then be reconstructed as 
\begin{align}
    \label{eq:demix-distributions}
    p_{Q}(x) &= \frac{p_{M_1}(x) - \kappa_{12}\,p_{M_2}(x)}{1-\kappa_{12}}\notag\\
    p_{G}(x) &= \frac{p_{M_2}(x) - \kappa_{21}\,p_{M_1}(x)}{1-\kappa_{21}}.
\end{align}
Importantly, the observable $x$ in the above reconstruction need not be the same as that which was used to extract the reducibility factors. This is because the mixture fractions are collective properties of the samples, not the individual jets. The jet topics framework has proven to be powerful for examining quarks and gluons separately, both from theoretical \cite{Komiske:2018vkc, Larkoski:2019nwj, Takacs:2021bpv, Stewart:2022ari} and experimental perspectives \cite{ATLAS:2019rqw, Brewer:2020och, Ying:2022jvy, Komiske:2022vxg, LeBlanc:2022bwd}.

In existing practical studies that use jet topic disentangling, the pure distributions of Eq.\,\ref{eq:demix-distributions} are constructed from histograms of the training mixtures. In principle one could use these histograms to estimate likelihoods according to the bin occupancy or generate samples using the inverse cumulative probability. However as one considers larger dimension, the computational cost\footnote{Both computation time and memory requirements scale with the number of histogram bins.} of these tasks increases and the density of the dataset within the space decays exponentially, which would lead to large errors. In addition to this curse of dimensionality, statistical errors are amplified when the bin values are subtracted during construction of the topics.

We propose to alleviate these issues with deep generative networks, which are able to model high-dimensional probability distributions including correlations. Since these models are unbinned, they have more efficient scaling to high dimension. Further, if an appropriate training procedure is used, one can learn the topic distributions directly from the quark/gluon mixtures, without the need for explicit subtraction. 
\subsection{Normalizing flows}

In this work, we model the topic distributions using normalizing flows (NFs)~\cite{RezendeNF,Kobyzev_2021,JMLR:v22:19-1028}. An NF is a density estimation model that aims to fit an unknown probability distribution from finite data samples. Being an unsupervised method, these models have gained popularity in collider physics for anomaly detection \cite{Nachman:2020lpy,Hallin:2021wme,Jawahar:2021vyu,Butter:2022lkf, Hallin:2022eoq}. Furthermore, a key property of NFs is that they are invertible, and so they can also be used as generative models. This mode of operation has also been explored in a particle physics context\,\cite{Gao:2020zvv,Choi:2020bnf,Lu:2020npg,Winterhalder:2021ave,Krause:2021ilc,Krause:2021wez,Butter:2021csz,Verheyen:2022tov,Kach:2022qnf, Kach:2022uzq}.

In essence, an NF represents a mapping $g:Z\to X$ from a known latent distribution $p_Z$ (often a gaussian) in a latent space $Z$,  to a distribution $q$ in the data space $X$ of the same dimension. If $g$ is invertible, then the density of a data point $x\in X$ is given by the change of variables formula,
\begin{equation}
    q(x) = p_Z(g^{-1}(x)) \left|\det \frac{\partial g^{-1}}{\partial x}\right|.
\end{equation}
The map $g$ is usually parametrized by a neural network with weights $\theta$. The network can be trained to model an arbitrary data distribution $p_X$ by minimizing the Kullback-Leibler (KL) divergence between $p_X$ and $q_\theta$. One can show that this is equivalent to minimizing the negative log-likelihood of data samples under the model:
\begin{align}
    \operatorname{KL}(p_X||q_\theta) &= \int\mathrm{d}x\,p_X(x)\log\left(\frac{p_X(x)}{q_\theta(x)}\right) \notag\\
    &= \left\langle\log\left(\frac{p_X(x)}{q_\theta(x)}\right)\right\rangle_{x\sim p_X} \notag\\[4pt]
    &= -\left\langle\log q_\theta(x)\right\rangle_{x\sim p_X} + \mathrm{const.}
\end{align}

Training against the above objective requires that one can sample directly from the target distribution. In the case of quark and gluon jets, we instead have access only to mixed datasets $M_i$. However, Eq.\,\ref{eq:demix-distributions} describes the pure distributions as linear combinations of these mixture distributions, which can be substituted into the equation above in order to split the integral. Taking the quark distribution as an example, this leads to
\begin{align}
    \label{eq:topic-flow-loss}
    \operatorname{KL}(p_Q||q_\theta)
    &= \kappa_{12}\left\langle\log q_\theta(x)\right\rangle_{x\sim p_{M_2}} - \left\langle\log q_\theta(x)\right\rangle_{x\sim p_{M_1}}\notag\\&\quad+ \mathrm{const.}\,,
\end{align}
and similarly for the gluon distribution. Thus we can use batches from the mixed datasets in order to learn the correct underlying components. Intuitively, the NF is trained to increase the likelihood of one mixture while decreasing the likelihood of the other. Since the only statistical difference between the datasets is their mixture fraction, the flow will learn the distribution of whichever class the first mixture is enriched with compared to the second.

This approach to training a normalizing flow is not limited to jet topics. Indeed, the expansion that lead to Eq.\,\ref{eq:topic-flow-loss} can be applied to any distribution defined as a linear combination. For example, this loss function can be used to train flows to perform event subtraction, which has been explored using generative adversarial networks (GANs)~\cite{Butter:2019eyo}.

\begin{figure*}
    \centering
    \includegraphics[width=0.95\textwidth]{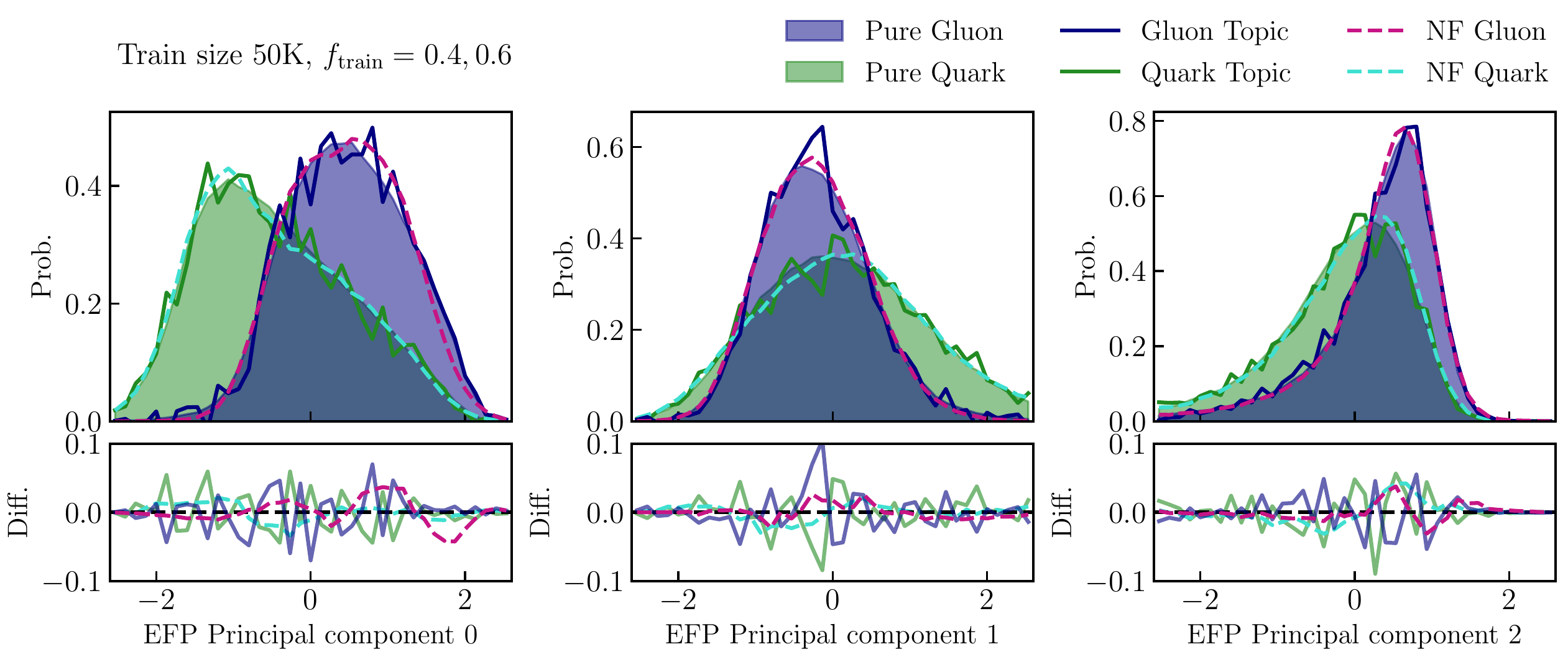}
    \includegraphics[width=0.95\textwidth]{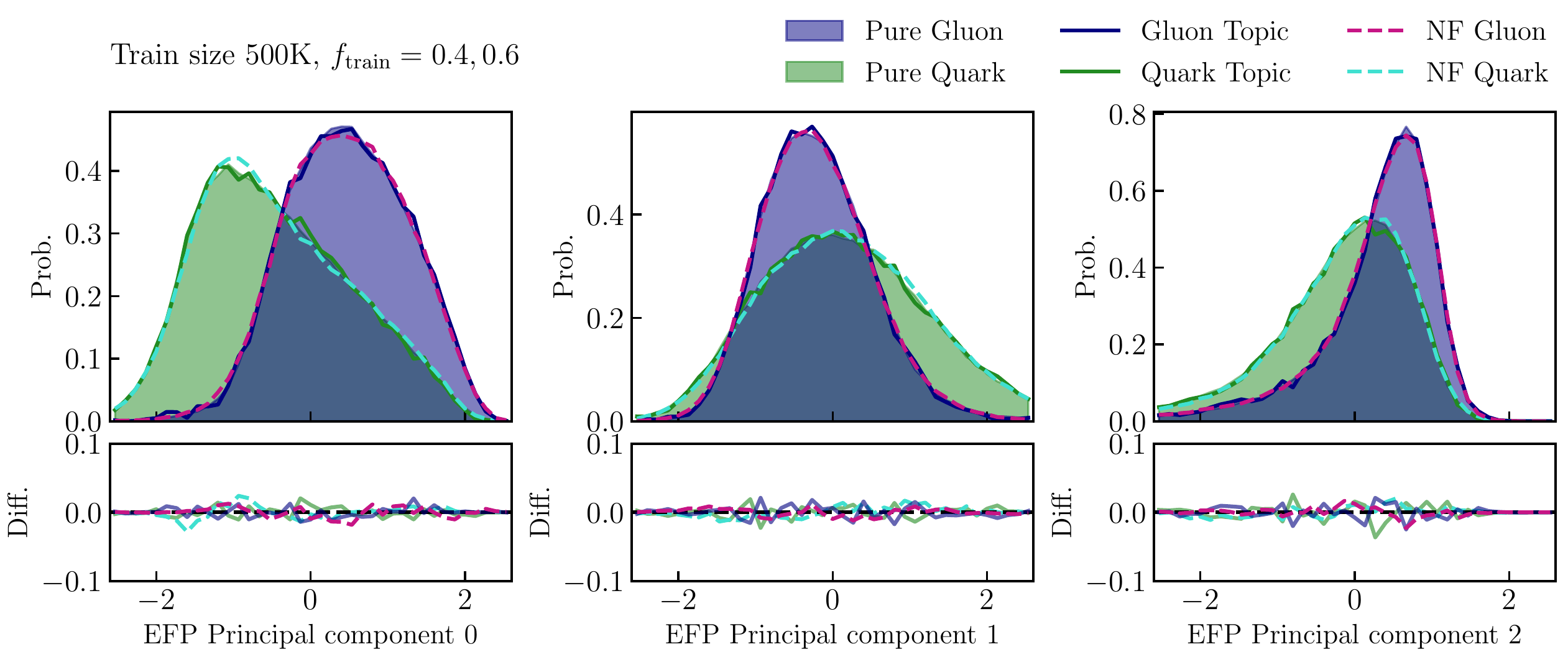}
    \caption{Validation of EFPs (3 of 8 components shown) generated by a normalizing flow trained on 40\% and 60\% quark mixtures. The shaded distributions are the pure quarks and gluons from the {\sc Pythia} test set and the solid lines are jet topic distributions constructed from the training set. The top and bottom rows are results for training set sizes of 50K and 500K respectively. Plots of the remaining components can be found in Appendix~\ref{app:plots}.}
    \label{fig:cond-agreement}
\end{figure*}

\section{Training details}
\label{sec:training}

\subsection{Datasets}
To emulate impure samples, we use the quark/gluon dataset of Refs.\,\cite{energyflow-quark-gluon, Komiske:2018cqr} which consists of 2M light quark~($uds$) and gluon jets generated with {\sc Pythia~8.2}~\cite{Sjostrand:2014zea}. We use 15\% for testing, 10\% for validation and up to 75\% for training. The jets come clustered by the anti-$k_T$ algorithm~\cite{Cacciari:2008gp} with radius $R=0.4$ and satisfy $p_T\in[500, 550]$ GeV and $|y|<1.7$.
Mixed datasets are constructed by collecting the appropriate number of quarks and gluons from each subset given the desired quark purity of each sample. As such, we have statistically-identical underlying quark/gluon distributions in all mixtures as required by the jet topics framework.\footnote{This condition is not necessarily satisfied in experimental data~\cite{Bright-Thonney:2018mxq, Komiske:2022vxg}.} We also have access to the exact reducibility factors of the mixtures. While in practice these would first need to be estimated using one of the methods described in Ref.\,\cite{Komiske:2022vxg}, we use the true values for simplicity.

In principle our procedure does not depend critically on the jet representation, so long as the representation can be integrated with a normalizing flow architecture. We choose to represent the jets as sets of Energy Flow Polynomials (EFPs)~\cite{Komiske:2017aww}, a complete basis of IRC-safe jet observables. An EFP can be identified with a multigraph where each node gives an energy-weighted sum over particles, and each edge denotes an opening angle between two particles. The polynomial corresponding to a particular graph $G$ with $N$ nodes is
\begin{equation}
	\mathrm{EFP}_G = \sum_{i_1=1}^{M} \cdots \sum_{i_N=1}^{M} z_{i_1} \cdots z_{i_N} \prod_{(j,k)\in G} \theta_{i_ji_k}^{\,\beta}\,,
\end{equation}
where $M$ is the number of jet constituents, $z_a=p_{T,a}/p_{T,J}$ is the $p_T$ fraction of constituent $a$ relative to the jet and $\beta$ is a parameter. EFPs are suitable for the present study since normalizing flows require a fixed-size representation. They are also strong quark/gluon discriminants. In fact a growing collection of recent work has demonstrated that, with careful selection, only a small number of EFPs are required to achieve classification on par with large neural networks based on low-level information~\cite{Faucett:2020vbu, Collado:2020ehf, Collado:2020fwm, Faucett:2022zie, Cal:2022fnm, Das:2022cjl}. As such, we opt to use a small set, namely the $\beta=\frac{1}{2}$ connected EFPs with ess than or equal to 3 edges, of which there are 8 elements.\footnote{We also exclude the single $d=0$ EFP under which all jets are degenerate.} Training in higher dimension is also possible, but requires larger networks in general (See Ref.\,\cite{Coccaro:2023vtb} for a dedicated study on NF scaling to high dimension as well as Refs.\,\cite{Krause:2021ilc,Krause:2021wez,Kach:2022qnf, Kach:2022uzq} for examples). That said, even in 8 dimensions, working with a histogram is impractical since just 10 bins per dimension leads to $10^8$ total bins.

To improve the stability of the models' training, we preprocess the EFPs by applying a log-scaling then translating the datasets to have zero mean (across quark and gluon jets). The flows also benefit from removing linear correlations in the data, so we additionally change basis to the principal components. The mean and covariance matrix are computed from the training split of each dataset. Since each of these preprocessing steps is invertible, the generated samples can always be converted back into the original EFP basis.

\subsection{Models}

We employ continuous-time normalizing flows~\cite{NEURIPS2018_69386f6b} as our generative models. Specifically, we use the {\tt tensorflow\_probability}~\cite{DBLP:journals/corr/abs-1711-10604} implementation of FFJORD~\cite{DBLP:conf/iclr/GrathwohlCBSD19}. The derivative network is a residual network with two blocks, each containing two layers of 256 nodes. We use tanh activations and minimize the loss of Eq.\,\ref{eq:topic-flow-loss} with the Adam optimizer. We set the batch size to either 1000 or the largest value that allows 25 batches per epoch, whichever is smaller. A callback that monitors the validation loss over each epoch of training is used to reduce the learning rate by a factor 10 if no improvement is observed after 5 epochs. We use an initial learning rate of $10^{-3}$ and halt training once the learning rate falls below $10^{-6}$.

\section{Results}
\label{sec:results}

\begin{figure*}[t]
    \centering
    \includegraphics[width=\textwidth]{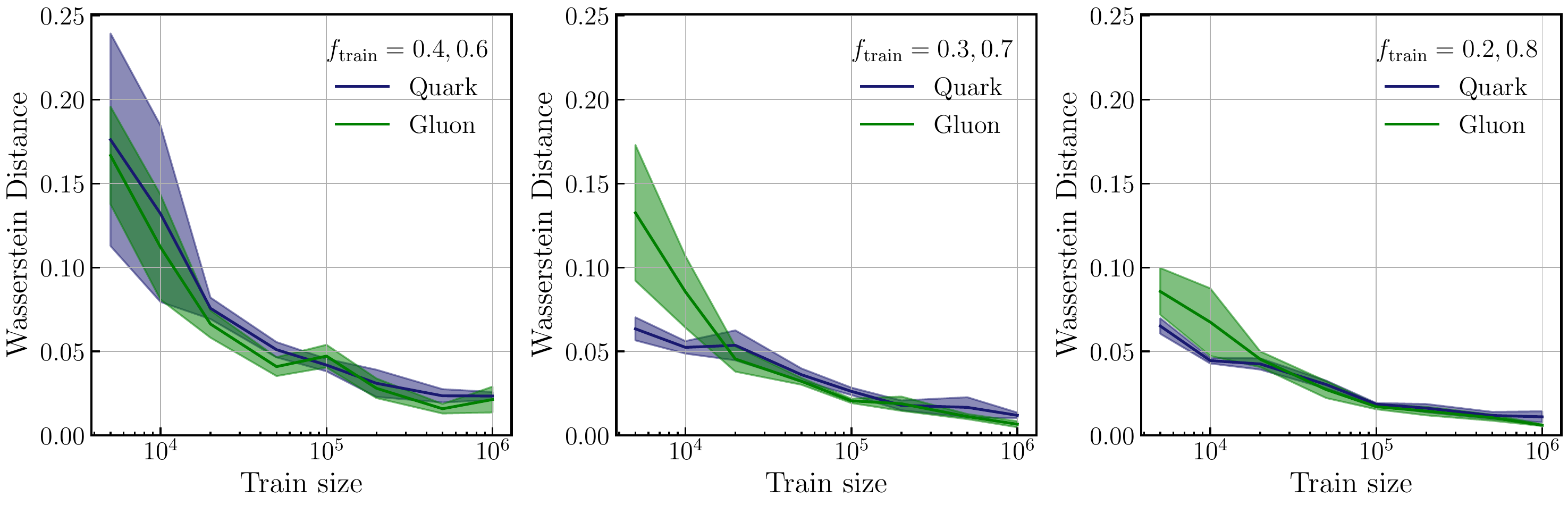}
    \caption{Average Wasserstein distances between EFP distributions in the {\sc Pythia} test set and those generated from networks trained on mixtures of different purity. Error bands show one standard deviation over 5 runs.}
    \label{fig:wasserstein}
\end{figure*}
\begin{figure*}[t]
    \centering
    \includegraphics[width=\textwidth]{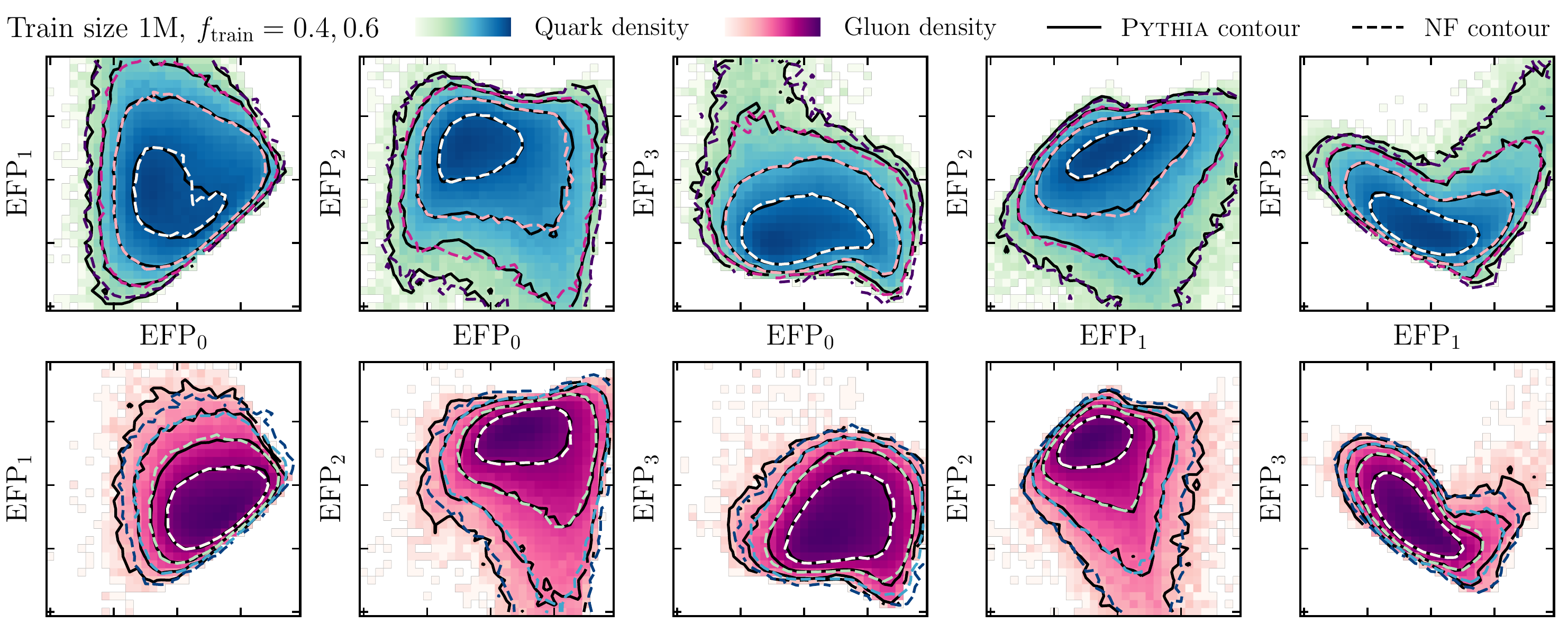}
    \caption{Comparison between the correlations amongst EFPs present in the {\sc Pythia} dataset and NF-generated samples (five of 28 pairs shown). The heat maps represent the simulated data density, with darker colors indicating larger density. The solid black lines are the contours of this density and the dashed colored contours map the density of generated samples. The top and bottom rows correspond to quarks and gluons respectively. Plots of the remaining pairs can be found in Appendix~\ref{app:plots}.}
    \label{fig:correlations}
\end{figure*}

\subsection{Sample quality}\label{subsec:samples}
We demonstrate the performance of the trained models in Fig.\,\ref{fig:cond-agreement}, where the pure EFP distributions from the {\sc Pythia} test set are compared with equal-sized samples generated with the flows. We show the first three components out of the eight which we use. We also compare to the jet topic distributions constructed from the training set. The plots show results for a small training set containing 50K jets (top row) as well as a large training set containing 500K (bottom row). The lower sub-panels in each plot show the absolute difference between the topic/NF and the pure distribution derived from \textsc{Pythia}. In the smaller training set, the histogram-based topics exhibit large statistical fluctuations which are not present in the NF distribution, since (a) no subtraction is performed and (b) the models have been sampled beyond the size of the training set. The flow samples instead exhibit systematic error corresponding to the fact that the networks cannot be trained to perfect convergence. The systematic variations of the flow appear to be equal or smaller than the statistical fluctuations of the binned topics. In the larger dataset, the statistical and systematic errors are diminished and both methods agree with the pure distributions derived from the truth-labels from \textsc{Pythia}. We find that the fractional errors for the 50K training sample are around 20\% and around 5\% for the 500K training sample near the peaks of the distributions for both standard topics and normalizing flows, although these become larger in the tails where statistical uncertainties are more important.

To concretely measure the quality of a trained NF, we evaluate 1-Wasserstein distances between the testing sample and an equal-size sample generated by the flow for each EFP in the set. Fig.\,\ref{fig:wasserstein} shows plots of the average of these distances against the size of the training set for mixtures with quark fractions $(0.4,0.6)$, $(0.3,0.7)$ and $(0.2,0.8)$.\footnote{We present results only for symmetric mixture purity (${f_2=1-f_1}$) to ensure that we have balanced training sets. This is not a limitation of the method in general, and we verified that results of equal quality are obtained using asymmetric purity.} As expected, the agreement between the test set and the NF samples improves with the size of the training set. For the most part, the quark and gluon distributions are reproduced equally well, however for small datasets, the flows do not perform as well for gluons. This is due to a tendency for the models to overfit in these settings, and so training was halted early. This could likely be addressed by optimizing the network size depending on the dataset at hand. One could also set the reducibility factor in Eq.\,\ref{eq:topic-flow-loss} to zero for some warm up period during training, effectively pretraining the model on one of the available mixtures.

The figure also reveals a dependence on the purity of the training mixtures. The NFs perform better when trained on mixtures with quark fractions away from 0.5, particularly when the dataset is small. This simply reflects the fact that optimizing the objective Eq.\,\ref{eq:topic-flow-loss} is more difficult when $M_1$ and $M_2$ are similar.

\begin{figure*}
    \centering
    \includegraphics[width=0.92\textwidth]{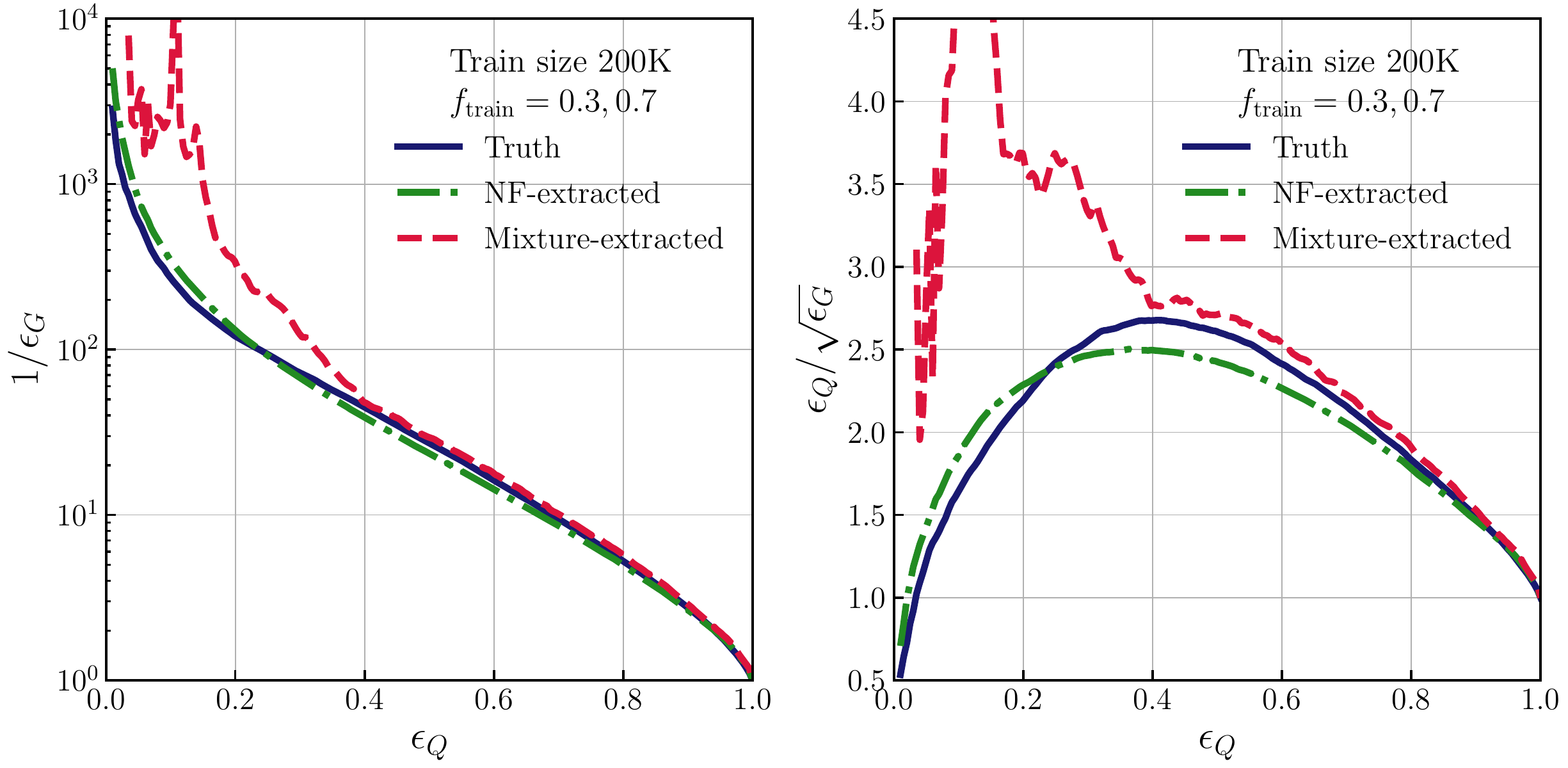}
    \caption{ROC (left) and SI (right) curves of a CWoLa classifier evaluated using  Eq.\,\ref{eq:efficiencies} applied to the training mixtures and using samples from NF-learned topic distributions.}
    \label{fig:calibration}
\end{figure*}

Since the NF models represent multi-dimensional probability distributions, correlations between EFPs should also be captured. To judge the pairwise correlations produced by the flow models we plot the quark and gluon distributions projected onto a selection of 2-dimensional planes in the full EFP space, shown in Fig.\,\ref{fig:correlations}. The heat maps and solid black contours correspond to the (truth) {\sc Pythia} distributions and the dashed contours are produced from equal-size samples of a trained NF. The non-trivial profiles of these distributions illustrate that correlations indeed remain between the components even after {rotating to the principal component basis}. The two sets of contours agree to a precise level, demonstrating that the generative model indeed produces the correct correlations. As a consequence, any cuts on a particular EFP in a generated sample will appropriately affect the distributions of the other EFPs without sacrificing statistics (since more samples can always be generated from the flow).

\subsection{Classifier calibration}

One of the advantages of an operational definition of quarks and gluons is that the performance of a classifier can be evaluated in a data-driven manner. As discussed in Ref.\,\cite{Komiske:2018vkc}, this enables weakly-supervised classifiers to self-calibrate using jet topics defined by their output. This proceeds by training a classifier on mixtures $M_1$ and $M_2$, then extracting reducibility factors and corresponding mixture fractions from the output $x$ using Eq.\,\ref{eq:red-factors}. Signal and background efficiencies can then be calculated for a given threshold $t$ on the output according to
\begin{equation}
    \epsilon_Q(t) = \frac{p_{M_1}(x>t)(1-f_2) - p_{M_2}(x>t)(1-f_1)}{f_1-f_2}\,, \notag
\end{equation}
\begin{equation}
    \label{eq:efficiencies}
    \epsilon_G(t) = \frac{p_{M_2}(x>t)f_1 - p_{M_1}(x>t)f_2}{f_1-f_2}\,.
\end{equation}
These efficiencies are again subject to amplified statistical uncertainty due to the subtraction. Thus errors can be quite large, especially in regions of low efficiency. However, one is often interested in regions of low background efficiency, especially for powerful classifiers. We find that even for sizeable datasets ($\sim10^5$ events) there can be insufficient statistics to produce smooth ROC curves throughout the entire region of interest. This is because larger datasets train stronger classifiers which increase class separation and thereby preserve sparsely occupied regions of the output space. This usually leads to a systematic underestimate of the background efficiency and can prevent reliable calibration.

As a concrete example, we train a CWoLa classifier~\cite{Metodiev:2017vrx} and plot the receiver operating characteristic (ROC) and significance improvement (SI) curves yielded by Eq.\,\ref{eq:efficiencies}, comparing with the \textsc{Pythia} truth. We show this in Fig.~\ref{fig:calibration} as the red dashed line. At high quark efficiencies, the mixture-extracted curves agree well with the truth. However, below an efficiency of 0.4 the curves diverge due to low statistics. Since the deviation causes the curve to reach its maximum significance improvement far from the true value, it is not an accurate means of calibration.

We can address this sensitivity to the training statistics with dedicated generative models for the pure distributions. Instead of using Eq.\,\ref{eq:efficiencies}, samples can be generated from each flow and passed through the classifier yielding the ``pure" class predictions.\footnote{One could also train an NF to disentangle the outputs themselves, although this would need to be done separately for each classifier.} These predictions can then be used to estimate efficiencies directly, without the need for any subtraction. We emphasize that oversampling classifier outputs in this way relies on full multidimensional generative models for the topic distributions since the classifier may be sensitive to high-order correlations.

The ROC and SI curves that we evaluate using the flow models are also shown in Fig.\,\ref{fig:calibration}. Compared to the mixture-extracted curves, those obtained from the NF samples remain close to the truth for a larger range of efficiencies. Critically, the location of the maximum significance improvement agrees closely with the truth. Once again, the trade-off is between statistical and systematic error, although we generally find that it is better to use the generative model.

\subsection{Generative classification}

\begin{figure}
    \centering
    \includegraphics[width=0.47\textwidth]{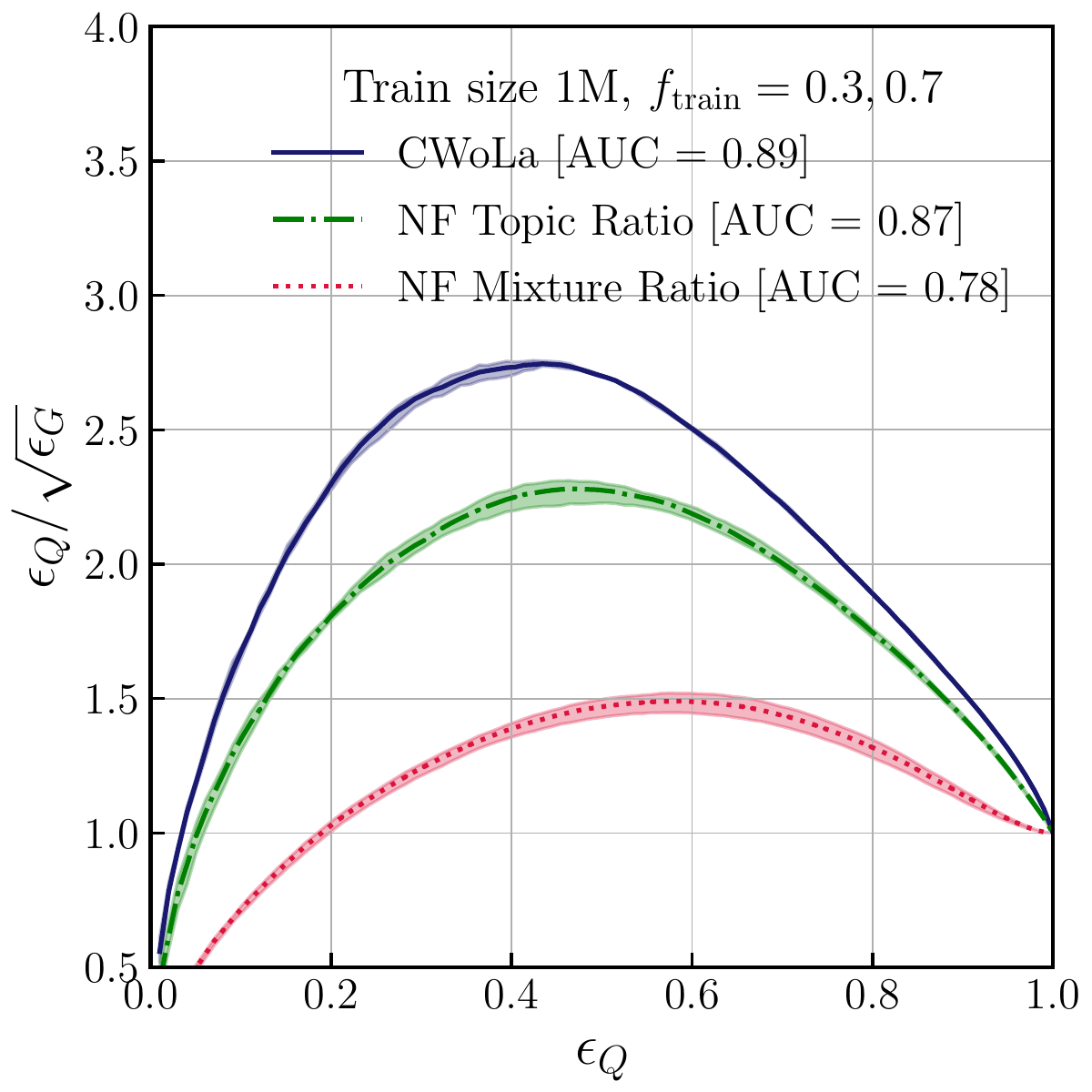}
    \caption{SI curves for a discriminative CWoLa classifier ($5\times100$ fully-connected layers) and two NF-based generative classifiers. The topic and mixture ratios are $p_Q(x)/p_G(x)$ and $p_{M_1}(x)/p_{M_2}(x)$ respectively. Efficiencies are evaluated using \textsc{Pythia} and error bands show the range of 5 independent classifiers.}
    \label{fig:generative-classification}
\end{figure}

In addition to sampling, another advantage of unbinned models for pure quark and gluon distributions is their ability to evaluate likelihoods of points in the full data space. In particular, one can use the trained normalizing flows to construct a quark/gluon classifier via the likelihood ratio $p_Q(x)/p_G(x)$.\footnote{In practice we use the log likelihood ratio.} Such a model is known as a generative classifier \cite{NIPS2001_7b7a53e2}. Compared with typical discriminative classifiers, these models are often not as performant since they model the likelihood functions globally, whereas only the decision boundary is relevant for classification. Indeed, we find that likelihood ratios constructed with our normalizing flows are unable to match a discriminative CWoLa classifier. However, using TopicFlow to learn the quark/gluon ratio leads to much stronger classification than using standard normalizing flows to construct the mixture ratio defined as $p_{M_1}(x)/p_{M_2}(x)$, despite the fact that the true ratios are related monotonically. Fig.\,\ref{fig:generative-classification}, shows the significance improvement curves for these generative classifiers, compared to a green discriminative CWoLa classifier.\footnote{For these results, we use a normalizing flow architecture consisting of coupling blocks with rational quadratic spline transformations. Our continuous-time normalizing flows yielded extremely poor classification.} The performance of the NF mixture ratio can be explained by the poor out-of-distribution sensitivity of normalizing flows~\cite{Kirichenko2020WhyNF}. Specifically, we find that gluons typically have higher likelihood than quarks even under an NF trained on a quark-enriched mixture. However, the positive term in Eq.\,\ref{eq:topic-flow-loss} explicitly penalizes this for TopicFlow, leading to better-separated likelihoods and thereby better classification.

While the raw performance of the quark/gluon NF ratio does not match CWoLa, generative classifiers offer another benefit. Specifically, they provide a means of detecting anomalous events that have small likelihoods under both quark and gluon models.
This sort of outlier detection is absent in discriminative models since their class predictions are normalized~\cite{NEURIPS2020_593906af, 9578744}. For example, a top-quark jet may have a small likelihood under both the light quark and gluon topic distributions, yet still be more compatible with the quark topic. In such a case, a discriminative classifier may confidently predict the jet to belong to the light quark distribution. However, in the generative case, one could use the small individual likelihoods to identify the jet as out-of-distribution.

This behavior is particularly desirable for application to real data, where the datasets are not guaranteed to contain only quark or gluon jets. By defining an anomaly score using individual likelihoods, the total significance improvement of a generative classifier may exceed the CWoLa tagger when including a cut on this score.

\section{Conclusions and Outlook}
\label{sec:conclusions}

We have introduced TopicFlow, a method of modeling jet topic distributions using normalizing flows. Our approach has two advantages over the standard procedure based on histograms. Firstly, the generative models can be oversampled to produce smooth distributions, avoiding amplified statistical uncertainty that arises when subtracting histograms. Secondly, the absence of binning allows the normalizing flows to be applied in high dimension and thus capture complex correlations in the data. Samples from the models are therefore suitable as input to other machine learning models. As an example, we demonstrated that our flows can enable self-calibration of a classifier that would otherwise be limited by a lack of statistics.

\begin{figure*}
    \centering
    \includegraphics[width=\textwidth]{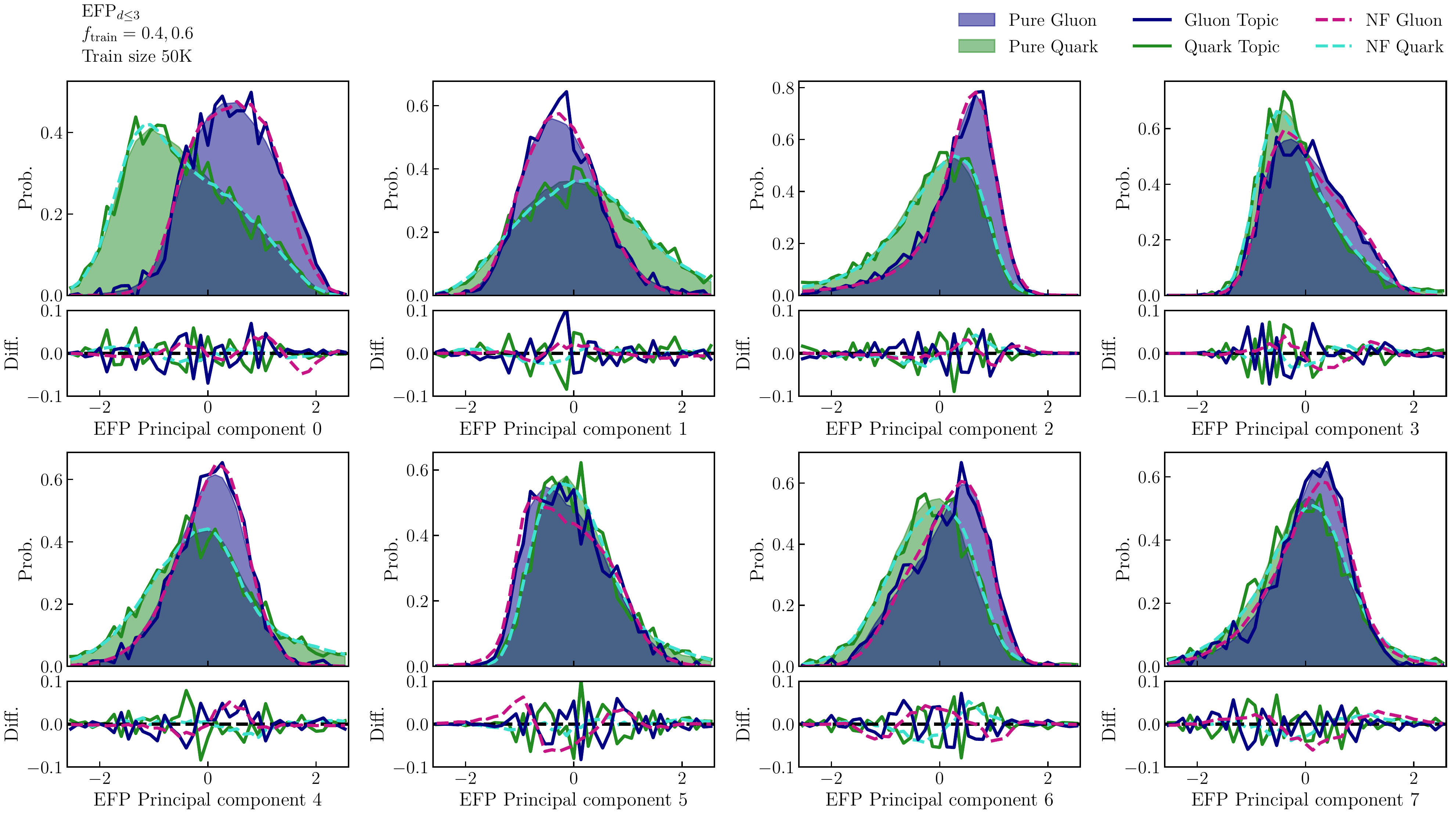}\\
    \vspace{1cm}
    \includegraphics[width=\textwidth]{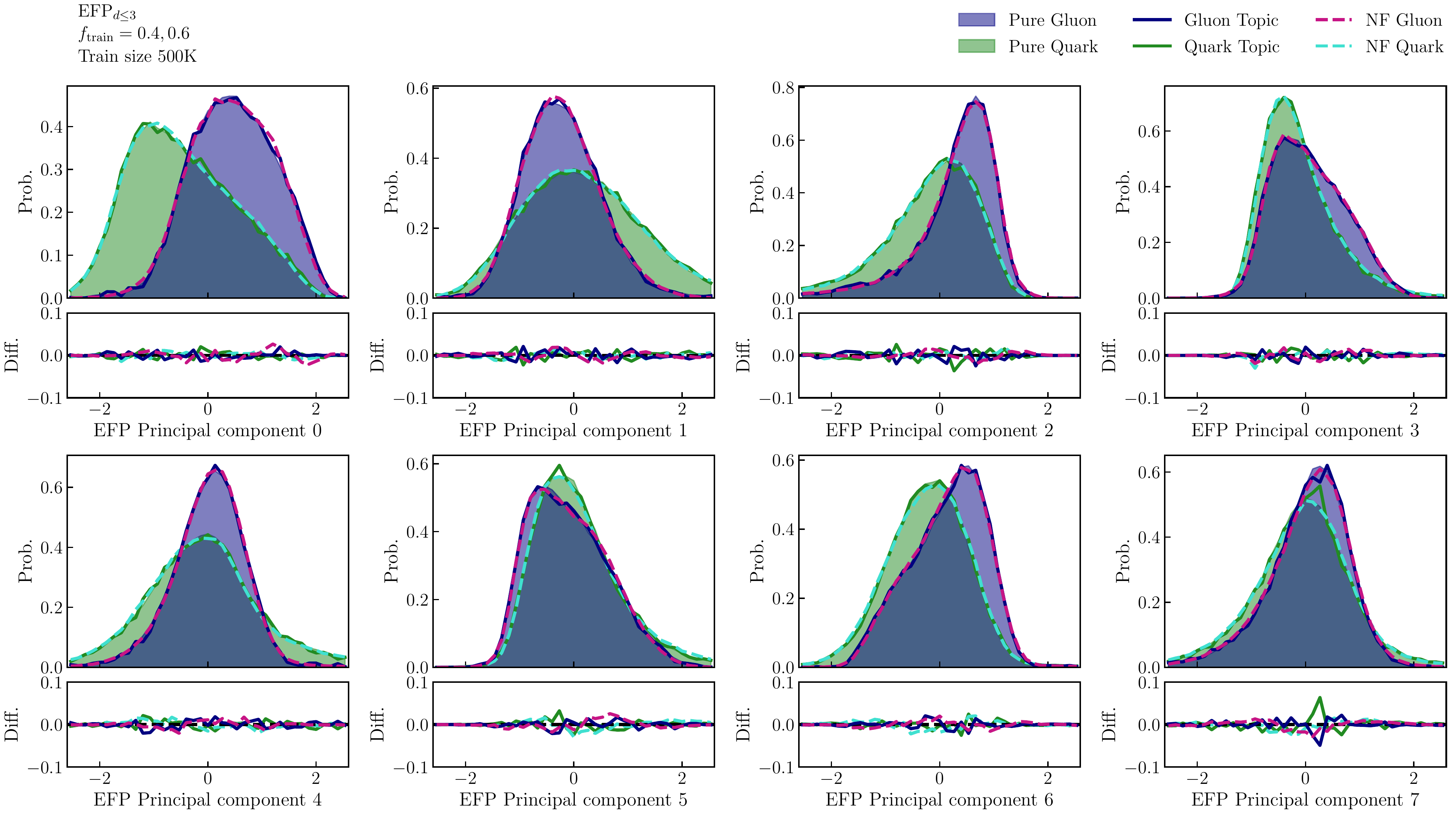}
    \caption{Comparison of quark and gluon EFP distributions as in Fig.\,\ref{fig:cond-agreement}, but including all principal components.}
    \label{fig:all-cond-agreement}
\end{figure*}

\begin{figure*}
    \centering
    \includegraphics[width=\textwidth]{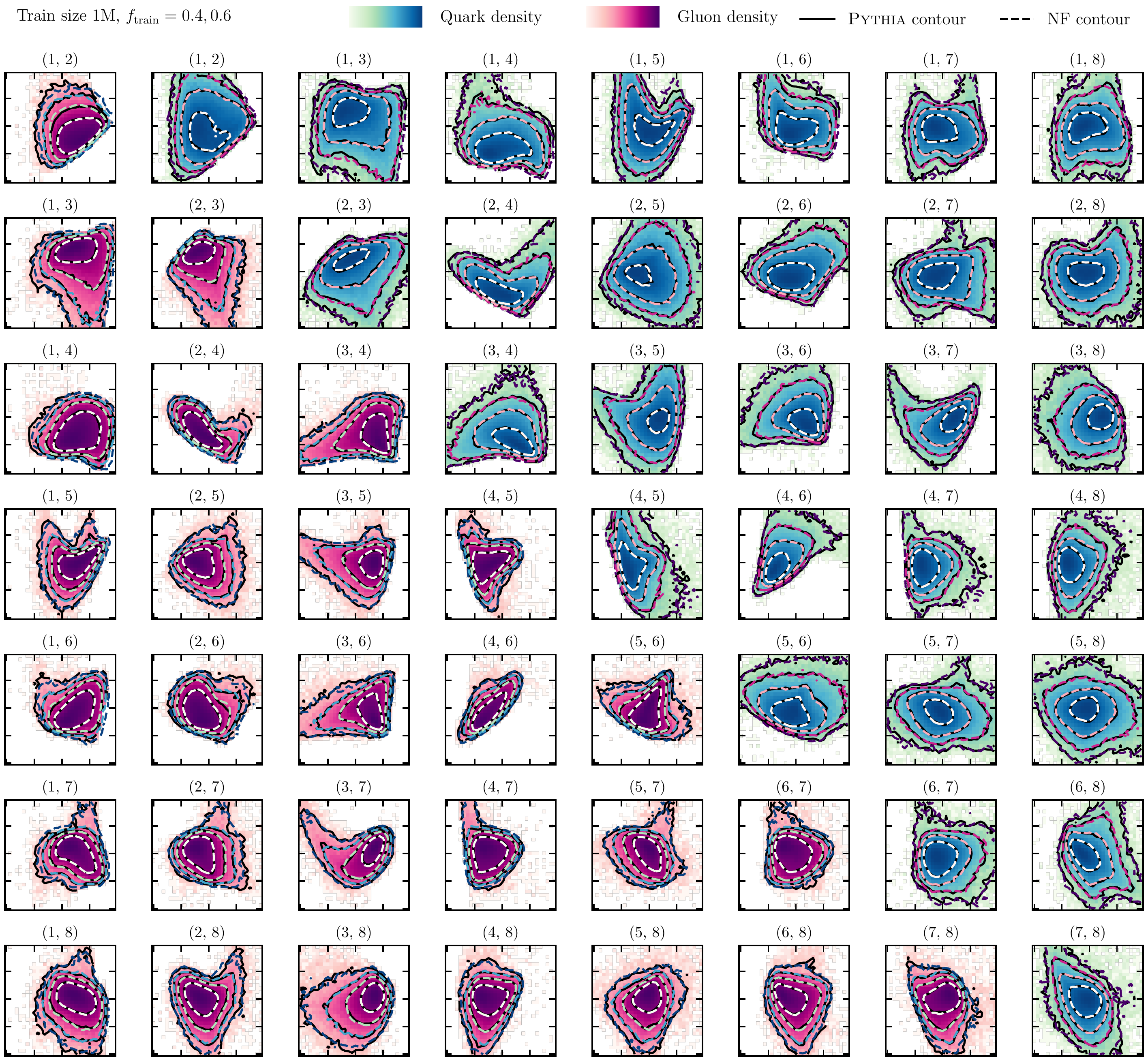}
    \caption{Comparison of pairwise EFP correlations as in Fig.\,\ref{fig:correlations}, but including all pairs between EFPs in the set.}
    \label{fig:all-correlations}
\end{figure*}

Normalizing flows allow the possibility of generative classification via use of the likelihood ratio. We found that using the ratio defined by TopicFlow is superior to using the ratio of mixture likelihoods with standard normalizing flows, though still not as performant a dedicated discriminative classifier based on the CWoLa method. However, classification using the individual likelihoods in this way may be more robust to out-of-distribution events in a testing sample and therefore warrants further study.

While the models in this work were trained on simulation with known quark fractions, the general framework can be extended to an unsupervised setting by first extracting mixture fractions/reducibility factors using existing methods such as~\cite{Komiske:2022vxg, Alvarez:2021zje}. As such a sensible direction for future work is to apply these networks to real jets in CMS Open Data.

It would also be interesting to explore particle-level jet representations such as point clouds. In this case, sampling allows for the determination of arbitrary jet observable distributions. Recent work has shown that normalizing flows can be trained effectively on such representations \cite{Kach:2022qnf, Kach:2022uzq}. GANs are also a natural choice for point cloud data \cite{Kansal:2021cqp} and could be applied to jet topics by training with the event subtraction method of Ref.\,\cite{Butter:2019eyo}. Conversely, one could use our training procedure to perform event subtraction with normalizing flows.

Other directions also include automatic extraction of the reducibility factors, uncertainty quantification with Bayesian neural networks and domain adaptation (in the jet $p_T$ for example) with conditional generative models. We hope to explore some of these in future work.

\section*{Code availability}
Our code is publicly available and can be found at \href{https://github.com/ayo-ore/topicflow}{https://github.com/ayo-ore/topicflow}.

\appendix
\section{Additional plots}
\label{app:plots}
Here we present extended versions of the plots in Section \ref{subsec:samples} that include all EFP dimensions. Fig. \,\ref{fig:all-cond-agreement} extends Fig.\,\ref{fig:cond-agreement} and Fig.\,\ref{fig:all-correlations} extends Fig.\,\ref{fig:correlations}.

\section*{Acknowledgements}
MJD is  supported by  the Australian Research Council Future Fellowship FT180100324. AO is supported by the Australian Government Research Training Program Scholarship initiative.
Computing resources were provided by the LIEF HPC-GPGPU Facility hosted at the
University of Melbourne. This Facility was established with the assistance of
LIEF Grant LE170100200.

\input{main.bbl}

\end{document}

%% file: main.bbl
%